\documentclass[10pt,letterpaper,twocolumn]{article} 

\usepackage{ol2}
\usepackage[draft]{hyperref}
\usepackage{amsmath}

\begin{document}

\twocolumn[ 

\title{Observing Abnormally Large Group Velocity at the Plasmonic Band Edge via a Universal Eigenvalue Analysis}


\author{Wei E.I. Sha,$^1$ Ling Ling Meng,$^1$ Wallace C.H. Choy,$^{1,*}$ and Weng Cho Chew$^{2,*}$}

\address{
$^1$Department of Electrical and Electronic Engineering, The University of Hong Kong,
Pokfulam Road, Hong Kong \\
$^2$Department of Electrical and Computer Engineering, University of Illinois, Urbana-Champaign,
Illinois 61801, USA \\
$^*$Corresponding authors: chchoy@eee.hku.hk (W.C.H. Choy); w-chew@uiuc.edu (W.C. Chew).
}

\begin{abstract}
We developed a novel universal eigenvalue analysis for 2D arbitrary nanostructures comprising dispersive and lossy materials. The complex dispersion relation (or complex Bloch band structure) of a metallic grating is rigorously calculated by the proposed algorithm with the finite-difference implementation. The abnormally large group velocity is observed at a plasmonic band edge with a large attenuation constant. Interestingly, we found the abnormal group velocity is caused by the leaky (radiation) loss not by metallic absorption (Ohmic) loss. The periodically modulated surface of the grating significantly modifies the original dispersion relation of the semi-infinite dielectric-metal structure and induces the extraordinarily large group velocity, which is different from the near-zero group velocity at photonic band edge. The work is fundamentally important to the design of plasmonic nanostructures.
\end{abstract}


 ] 

\noindent For photonic crystals \cite{1}, forward and backward traveling waves constructively interfere with each other, resulting in standing waves. Due to Bragg scattering in periodic photonic structures with modulated refractive indices, the degenerate standing waves split into two band edge modes with fields concentrated in different regions. Between band edge frequencies, a band gap occurs due to destructive interferences of traveling waves. Although it is nonintuitive that the phenomena could happen in metal structures, no physical law forbids their existences. Different from traveling waves in photonic crystals, forward and backward surface plasmon (SP) waves could also interfere with each other, leading to the formation of plasmonic band gap (PBG) and plasmonic band edge (PBE) \cite{2}. They have broad applications in nanophotonics such as thin-film solar cells \cite{3}, lasers \cite{4,5}, surface-enhanced Raman scattering \cite{6,7}, directional nanoantennas \cite{8}, etc.

The complex dispersion relation (or complex Bloch band structure) of metallic nanostructures plays a fundamental role in understanding the PBE and PBG effects. In literatures, the dispersion relation has been approximately generated by the frequency-angle diagram \cite{6,9}, which is not accurate for highly lossy metal materials. Because the eigenvalue (Bloch wave number) corresponding to the incident angle will be complex given a real frequency. In this work, we developed a novel and rigorous eigenvalue analysis for 2D arbitrary nanostructures with dispersive and lossy materials. Previously, the complex band structure can be calculated by converting the quadratic eigenvalue problem to linear one \cite{10,11,11_add}. Here we borrow an idea from the quantum transport problem \cite{12} to overcome the drawback. First, the proposed method can handle universal dispersive and lossy media with experimentally-tabulated complex refractive indices. Second, the method can produce a linear eigenvalue equation and thus saves computer resources in contrast to the quadratic eigenvalue one. Third, it is easy to be implemented without decompositions of electromagnetic fields. Finally, the proposed method can be employed to study the eigenvalue problem of arbitrarily shaped plasmonic nanostructures, which shows advantages over mode matching approaches \cite{12_add1,12_add2,12_add3} existing the numerical convergence issue.

Using the proposed eigenvalue algorithm, we observed an abnormally large group velocity at the PBE of a 2D metallic grating with 1D periodicity. The extraordinary group velocity distinguishes from the near-zero group velocity investigated both in photonic crystal structures \cite{1} and in other periodic plasmonic structures with untouched or separated unit elements \cite{2,5}.

\begin{figure}[!tbc]
\centering
\includegraphics[width=3.3in]{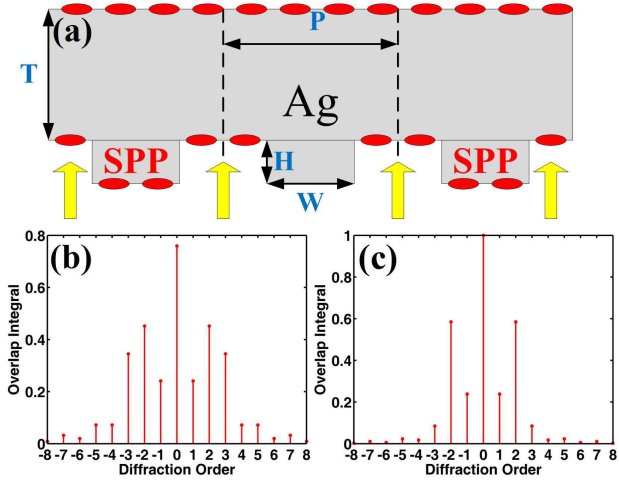}
\caption{(a) The schematic pattern of a squarely modulated metallic grating with the geometric sizes of $W=410$ nm, $P=820$ nm, $H=30$ nm, and $T=50$ nm. The red ellipses and yellow arrows denote the excited surface plasmons and incident light, respectively. (b,c) The normalized spatial overlap integrals between the near-field profile of the grating and the m-th Floquet mode respectively at $410$ nm (Peak 1 of Fig. \ref{fig2}(a)) and $480$ nm (Peak 3 of Fig. \ref{fig2}(a)).}
\label{fig1}
\end{figure}

\begin{figure}[!tbc]
\centering
\includegraphics[width=3.3in]{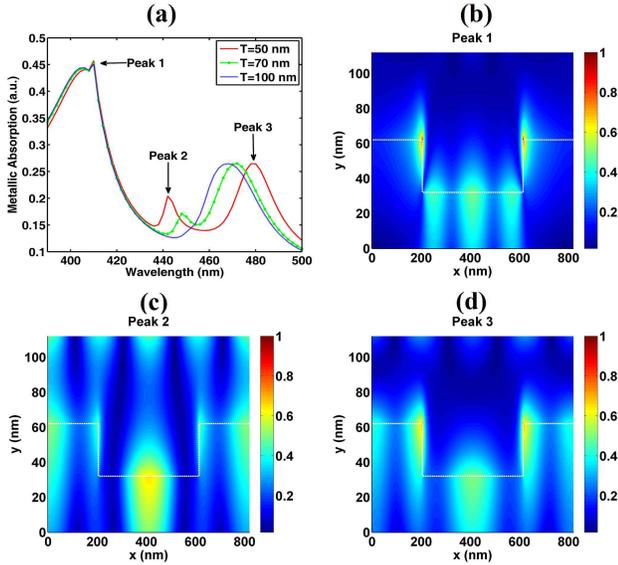}
\caption{(a) The optical absorption of the metallic grating with a varying thickness $T$. (b,c,d) $H_z$ field distributions of the metallic grating corresponding to the absorption peaks denoted by arrows ($T=50$ nm).}
\label{fig2}
\end{figure}

\begin{figure}[!tbc]
\centering
\includegraphics[width=3.3in]{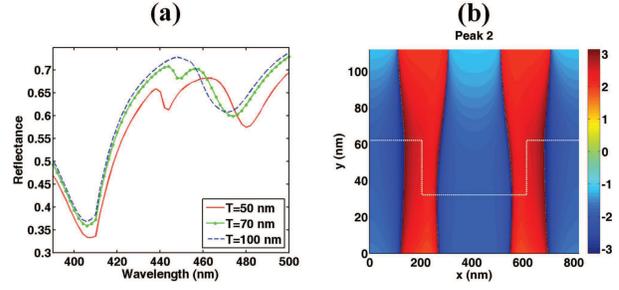}
\caption{(a) The reflectance of the metallic grating with a varying thickness $T$. (b) The phase distribution of $H_z$ field at the PBG with respect to the Peak 2 in Fig. \ref{fig2}(a).}
\label{fig2_add}
\end{figure}

Fig. \ref{fig1}(a) shows the schematic pattern of a squarely modulated metallic grating. SP waves will be excited at the crest and trough of the grating. The incident light is \textit{p} polarized and vertically impinges on the grating. The complex refractive index of the metal (Ag) is expressed by the Brendel-Bormann (B-B) model \cite{13}. The mismatched momentum is provided by Floquet modes of the periodic grating, which play an important role in inducing the PBE. Fig. \ref{fig2}(a) illustrates the optical absorption of the metallic grating. The middle absorption peak is related to the PBG, while others are related to PBEs. The weaker (middle) absorption peak by the PBG is due to the coupling between top and bottom SP waves as shown in Fig. \ref{fig2}(c). When the thickness of the grating $T$ increases, the middle absorption peak disappears and thus an absorption dip is formed by the PBG (See Fig. \ref{fig2}(a)) where the reflectance has a high peak (See Fig. \ref{fig2_add}(a)). Fig. \ref{fig2_add}(b) depicts the phase distribution of $H_z$ field at the PBG. We see a plasmonic standing wave pattern corresponding to the second-order Floquet mode. While the planar semi-infinite air-Ag structure only supports one plasmonic resonance peak that is significantly smaller than $400$ nm, the two resonant peaks of PBEs at $410$ nm and $480$ nm are obtained after introducing the grating structure. Intuitively, the original dispersion relation of the semi-infinite air-Ag structure is modified by the periodically modulated surface (refractive indices) of the metallic grating.

To clarify the modified dispersion relation, we theoretically study the spatial overlap integral between the near-field profile of the grating as shown in Figs. \ref{fig2}(b,d) and the m-th Floquet mode according to the rigorous coupled-wave analysis \cite{14}, i.e.
\begin{equation}
\eta=\left|\int_0^P\int_{0}^{y_c}H_z(x,y)\exp\left(jk_xx+jk_y(y-y_c)\right)dxdy\right|
\end{equation}
\begin{equation}
k_x=k_0\sin\theta+2\pi n/P,\,\,k_x^2+k_y^2=k_0^2,\,\,\mathrm{Im}(k_y)<0
\end{equation}
where $P$ is the periodicity of the grating, and $y_c$ is the boundary between the grating crest and air. Moreover, $k_0$ is the wave number in free space, $\theta=0$ for the vertical incidence case, and $n=0,\pm 1,\pm 2$ are integers. From a simple calculation ($\mathrm{Re}(k_0)>2\pi/P$), we know the zeroth- and first-order Floquet modes are radiative modes related to incident wave, specular reflection, etc. They are not responsible for the excitation of SPs. Furthermore, the overlap integral values become very small when the diffraction order $|n|>3$ (See Figs. \ref{fig1}(b,c)). Remarkably, the overlap integral value with respect to the third-order Floquet mode suddenly increases at $410$ nm as comparing Fig. \ref{fig1}(b) to Fig. \ref{fig1}(c). This also can be observed from the near-field distributions as depicted in Figs. \ref{fig2}(b,d), where the H-field is concentrated on the crest and trough of the grating, respectively. The interplay of second- and third-order Floquet modes not only enables a coupling of SPs to photon energy but also perturbs the original dispersion relation of SPs supported in the semi-infinite air-Ag structure. A proper momentum matching condition can be revised as
\begin{equation}
k_{sp}+\Delta k_{sp}=k_0\sin\theta+\frac{2\pi}{P}n
\end{equation}
where $k_{sp}=k_0\sqrt{\frac{\epsilon_0\epsilon_m}{\epsilon_m+\epsilon_0}}$, $\epsilon_0$ and $\epsilon_m$ are permittivities of free space and metal, $\Delta k_{sp}$ is the perturbed momentum by the third-order Floquet mode, and $n = 2$ corresponds to the second-order Floquet mode.

Besides the overlap integral method, the eigenvalue analysis is an efficient tool to characterize the PBE and PBG. Take 1D periodicity for an example, the wave equation can be discretized by using the finite-difference method and Bloch theorem $\phi(x+P)=\phi(x)\exp(-jk_BP)$, i.e.
\begin{equation}\label{eq4}
\begin{pmatrix}
  D_{11} & T & 0 & Te^{jk_BP} \\
  T & D_{22} & T & 0 \\
  0 & T & D_{33} & T \\
  Te^{-jk_BP} & 0 & T & D_{44} \\
\end{pmatrix}
\begin{pmatrix}
  \phi_1\\
  \phi_2\\
  \phi_3\\
  \phi_4\\
\end{pmatrix}=0
\end{equation}
where $\phi=[\phi_1;\phi_2;\phi_3;\phi_4]$ is the eigenmode, $D_{ii}=-2/\Delta^2+\left[k(x)\right]^2$, $T=1/\Delta^2$, and $\Delta$ is the spatial step. We define two new matrices $H$ and $Q$ as
\begin{equation}\label{eq5}
H=\begin{pmatrix}
  D_{11} & T & 0 & 0 \\
  T & D_{22} & T & 0 \\
  0 & T & D_{33} & T \\
  T & 0 & 0 & 0 \\
\end{pmatrix},\,
Q=-\begin{pmatrix}
  0 & 0 & 0 & T \\
  0 & 0 & 0 & 0 \\
  0 & 0 & 0 & 0 \\
  0 & 0 & T & D_{44} \\
\end{pmatrix}
\end{equation}
and use them to rewrite Eq. \eqref{eq4} as
\begin{equation}\label{eq6}
H\phi=e^{jk_BP}Q\phi
\end{equation}
Because $Q$ containing columns filled with zeros cannot be inverted and $H$ removing a diagonal element $D_{44}$ is singular, we should modify Eq. \eqref{eq6} to generate a well-conditioned linear eigenvalue problem
\begin{equation}\label{eq7}
\begin{split}
H\phi&=e^{jk_BP}Q\phi+Q\phi-Q\phi,\\
(H-Q)\phi&=(e^{jk_BP}-1)Q\phi,\\
(H-Q)^{-1}Q\phi&=\frac{1}{e^{jk_BP}-1}\phi,\\
M\phi&=\lambda\phi.
\end{split}
\end{equation}
The matrix $H-Q$ restores a standard finite-difference matrix of the wave equation and therefore can always be inverted. The above procedure can be easily extended to calculate the dispersion relation of the 2D metallic grating with 1D periodicity. The stretched-coordinate perfectly matched layers \cite{15} at the top and bottom boundaries are adopted to absorb outgoing waves reflected by the metallic grating. At the left and right boundaries, the same Bloch theorem is employed as Eq. \eqref{eq4}. It is worth mentioning the dispersion analysis of 2D gratings is an unbounded problem, which is fundamentally different from 2D periodic crystal structures with bounded boundaries.

\begin{figure}[!tbc]
\centering
\includegraphics[width=3.1in]{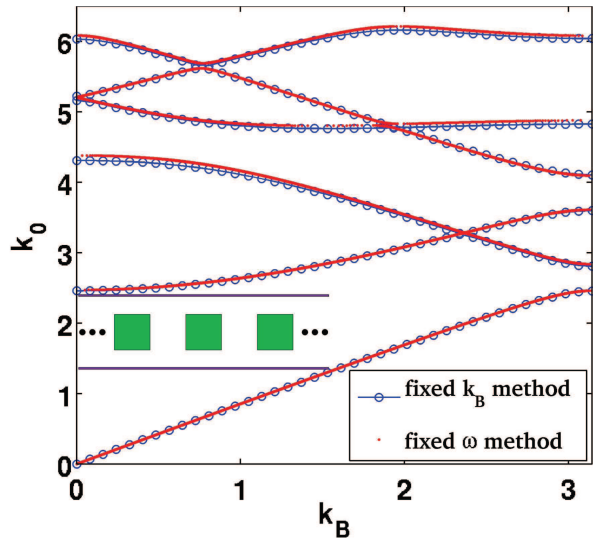}
\caption{The band structure of a periodic dielectric strip shielded with a perfect electric conductor plate. The dielectric constant and the side length of each strip is 4 and P/2, respectively.}
\label{fig3}
\end{figure}

The most commonly used method \cite{1} for calculating dispersion curves is to choose the Bloch wave number $k_B$ beforehand, and the frequency is then computed as a function of the wave number, yielding the dispersion curve $\omega=\omega(k_B)$. However, this traditional method is difficult to calculate the dispersion curve for the dispersive material. In this paper, we fix the frequency $\omega$ and solve for the Bloch wave number as a function of frequency, i.e. $k_B=k_B(\omega)$. To benchmark the proposed method, we compute the band diagram of a periodic dielectric strip shielded with a perfect electric conductor (PEC) plate. Both methods agree with each other very well, which can be seen in Fig. \ref{fig3}.

\begin{figure}[!tbc]
\centering
\includegraphics[width=3.1in]{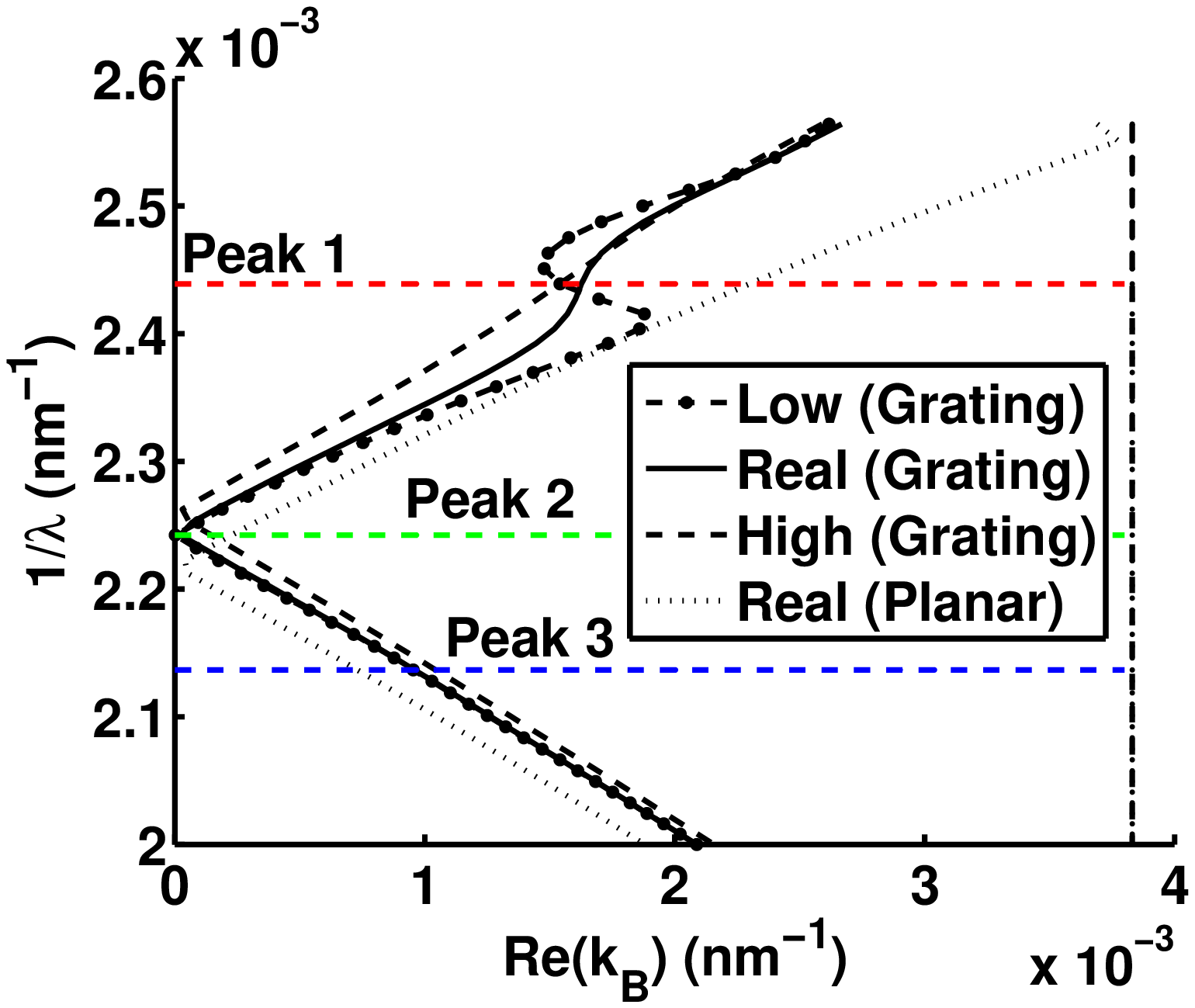}
\caption{The band structure of the 2D metallic grating as a function of the real part of the Bloch wave number. $\lambda$ is the incident wavelength and the right-most vertical line denotes $\mathrm{Re}(k_B)=\pi/P$. The imaginary part of (physically real) metal permittivity by the B-B model increases or decreases by a factor of 2 respectively for simulating the high or low Ohmic loss.}
\label{fig4}
\end{figure}

\begin{figure}[!tbc]
\centering
\includegraphics[width=3.1in]{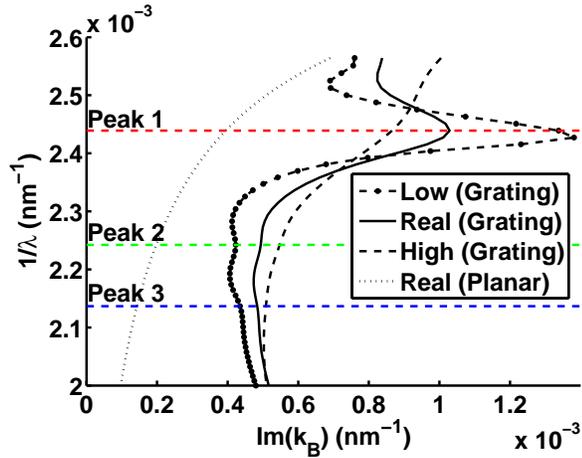}
\caption{The band structure of the 2D metallic grating as a function of the imaginary part of the Bloch wave number. $\lambda$ is the incident wavelength. The imaginary part of (physically real) metal permittivity by the B-B model increases or decreases by a factor of 2 respectively for simulating the high or low Ohmic loss.}
\label{fig5}
\end{figure}

Next, we will investigate the complex Bloch band structure of a 2D metallic grating as shown in Figs. \ref{fig4} and \ref{fig5}. At the PBE (denoted by the Peak 1), the original dispersion relation of the semi-infinite air-Ag structure is modified after the introduction of the periodically modulated refractive indices as illustrated in Fig. \ref{fig4}. Moreover, we find an extraordinarily large group velocity at the PBE. From Fig. \ref{fig5}, a large attenuation coefficient is found at the PBE corresponding to the large imaginary part of the Bloch wave number. To unveil the physical origin of the anomalous group velocity, we modify the imaginary part of metal permittivity. Interestingly, the negative and normal group velocity are observed respectively for the low and high metallic absorption (Ohmic) loss (See Fig. \ref{fig4}). Hence, the leaky (radiation) loss leads to the extraordinary group velocity while the Ohmic loss smoothens the plasmonic band structure losing distinct features. The PBE, also called PBG edge or edge of PBG in literatures \cite{8,16}, is spectrally located far from the PBG in contrast to photonic band edge where the group velocity is almost zero. The interplay of second-order and third-order Floquet modes in the grating strongly perturbs the original dispersion relation of the semi-infinite planar structure. The single modulated surface of the grating not only induces PBEs but also produces large leaky loss, which differs from plasmonic Bragg waveguides \cite{12_add1} where optical energy is confined between two modulated metal interfaces.

The PBG at the peak 2 with a small attenuation coefficient (See Fig. \ref{fig5}) is result from two counteracting SP waves with different group velocities (See Fig. \ref{fig4}). Compared to PBEs with significant plasmonic hot spots at grating corners, the PBG has a weaker optical absorption as depicted in Fig. \ref{fig2}(a). In other words, the PBG suppresses the Ohmic loss. The PBG is located at $\mathrm{Re}(k_B)=0$ that is the edge of irreducible Brillouin zone similar to the photonic band gap. It should be noted that the zero PBG is caused by the symmetric property of metallic gratings and the PBG could open up in dual-interface gratings \cite{16}.

The work proposed a novel universal eigenvalue solution to investigate the dispersion relation of a metallic grating. The extraordinary group velocity at the PBE is confirmed by our eigenvalue solver. The interaction between Floquet modes supported by the periodically modulated surface induces the strong leaky loss and thus the abnormally large group velocity. Both the developed numerical algorithm and new findings with detailed physical explanations are useful for designing plasmonic nanostructures.

The authors acknowledge the support of the grants (Nos. 712010, 711609, and 711511) from the Research Grant Council of the Hong Kong, the grant (No. 10401466) from the University Grant Council (UGC) of the University of Hong Kong (HKU), and the Seed Project Funding of HKU (No. 201211159147). This project is also supported by the UGC of Hong Kong (No. AoE/P-04/08), by The National Natural Science Foundation of China (No. 61201122), and in part by a Hong Kong UGC Special Equipment Grant (SEG HKU09).

\end{document}